\documentclass[10pt,aps,twocolumn,prc,floatfix,preprintnumbers,superscriptaddress,notitlepage,nofootinbib]{revtex4-2}
\usepackage[T1]{fontenc}
\usepackage{bm}
\usepackage{graphicx}
\usepackage{wrapfig}
\usepackage{array} 
\usepackage{listings}
\usepackage[para,online,flushleft]{threeparttablex}
\usepackage{xcolor}
\usepackage{mathrsfs}
\usepackage{multirow}
\usepackage{slashed}
\usepackage{amsmath}
\usepackage{booktabs}

\begin{document}

\title{Medium-mass nuclei with neural quantum states}

\author{Bryce Fore}
\affiliation{Physics Division, Argonne National Laboratory, Argonne, Illinois 60439, USA}

\author{Jane Kim}
\affiliation{Physics Division, Argonne National Laboratory, Argonne, Illinois 60439, USA}

\author{Alessandro Lovato}
\email{lovato@anl.gov}
\affiliation{Physics Division, Argonne National Laboratory, Argonne, Illinois 60439, USA}
\affiliation{Computational Science Division, Argonne National Laboratory, Argonne, Illinois 60439, USA}
\affiliation{INFN-TIFPA Trento Institute of Fundamental Physics and Applications, 38123 Trento, Italy}

\author{Anthony Tropiano}
\affiliation{Physics Division, Argonne National Laboratory, Argonne, Illinois 60439, USA}

\begin{abstract}
We compute ground-state energies and charge radii of light- to medium-mass nuclei with up to $A=58$ nucleons, leveraging a variational Monte Carlo method based on Pfaffian-Jastrow neural quantum states. To further understand which elements of the nuclear Hamiltonian are ``essential'' to predict binding energies and charge radii across the nuclear chart with few-percent errors, we consider different interactions inspired by pionless effective field theory. Specifically, in addition to model ``o'' of Ref.~\cite{Schiavilla:2021dun}, we study the impact of charge-symmetry-breaking and charge-dependent terms in the nucleon-nucleon force, as well as $p$-wave contributions, which have been found to be critical for the stability of $p$-shell nuclei. In addition to its intrinsic interest, our work assesses the performance of neural quantum states in the medium-mass regime and examines the impact of these interaction modifications. Using the resulting ground-state simulations, we analyze the computational scaling of variational Monte Carlo with neural quantum states as a function of system size and computational resources, enabling projections for future large-scale calculations.

\end{abstract}

\maketitle

\section{Introduction}
The accurate description of medium-mass nuclei remains a central challenge in nuclear theory, providing critical insights into the structure of nuclear forces, including the role of three-nucleon interactions~\cite{Otsuka:2009cs,Hagen:2012sh}. These nuclei are also the subject of intense experimental activity, ranging from rare-isotope-beam studies to electron-scattering experiments~\cite{Crawford:2023txq,Frois:1987hk,CREX:2022kgg}. Investigating these systems requires quantum many-body methods that solve the nuclear Schr\"{o}dinger equation with controlled approximations~\cite{Hergert:2020bxy}.

Several \emph{ab initio} approaches~\cite{Hergert:2020bxy}, including the in-medium similarity renormalization group~\cite{Tsukiyama:2010rj,Hu:2021trw}, coupled-cluster theory~\cite{Hagen:2008iw,Bonaiti:2025bsb}, and self-consistent Green's function methods~\cite{Soma:2019bso}, have achieved remarkable success in this region and can reach nuclei beyond medium mass, starting from interactions derived within chiral effective field theory. These methods, however, are commonly formulated in truncated single-particle basis expansions, often based on harmonic-oscillator orbitals. As a result, resolving correlations across different length scales may require large model spaces~\cite{Coon:2012ab,More:2013rma,Caprio:2012rv}, particularly for short-range correlations, extended asymptotic behavior, and cluster-like configurations~\cite{Kravvaris:2017nyj,Otsuka:2022bcf}.

Quantum Monte Carlo (QMC) approaches, both in the continuum~\cite{Carlson:2014vla} and on the lattice~\cite{Epelbaum:2011md}, naturally capture long-range correlations and clustering effects. In addition, continuum QMC methods excel at describing correlations across all relevant length scales, including the short-range correlations induced by the nuclear interaction when nucleon wave functions overlap. State-of-the-art implementations, such as Green's Function Monte Carlo (GFMC) and Auxiliary-Field Diffusion Monte Carlo (AFDMC), provide highly accurate benchmarks for light nuclei~\cite{Piarulli:2017dwd,Martin:2023dhl}. However, their extension to heavier systems remains challenging. GFMC is not applicable in this regime because of the exponential growth of the Hilbert space, while the accuracy of AFDMC deteriorates with increasing system size as a consequence of the fermion-sign problem and the use of variational wave functions that are not size extensive~\cite{Gandolfi:2014ewa}.

To address these challenges, neural quantum states (NQS)~\cite{Carleo:2016svm}, in which variational wave functions are parameterized by neural networks, have been applied to the nuclear many-body problem. Originally developed in the context of condensed matter and quantum chemistry, NQS have been combined with variational Monte Carlo (VMC) techniques over the past five years to study atomic nuclei~\cite{Keeble:2019bkv,Adams:2020aax,Lovato:2022tjh,Yang:2022esu} and nuclear matter~\cite{Fore:2022ljl,Fore:2024exa}. Their flexible representational power, combined with scalable optimization and the absence of a fermion-sign problem at the variational level, makes them a promising alternative for treating strongly correlated nuclear systems, with the potential to match or surpass the accuracy of GFMC and AFDMC methods. Beyond ground-state properties, NQS also enable the computation of dynamical observables, such as the low-energy photoabsorption cross section~\cite{Parnes:2025seu} and $\alpha$ scattering~\cite{Yang:2025mhg}. Finally, the description of hypernuclei has recently become possible~\cite{DiDonna:2025oqf,Zhang:2025okd}, including their excited states~\cite{Zhang:2026iex}.

Extending the analyses of Refs.~\cite{Lu:2018bat,Gnech:2023prs}, in this work we aim to elucidate which elements of nuclear Hamiltonians are essential to reproduce properties of nuclei across the nuclear chart. An open question is how sensitive bulk observables are to physics above the pion scale, and to what extent nuclear binding is governed by near-unitary nucleon-nucleon interactions. To address this question, we compute ground-state energies and charge radii for nuclei with up to $A=58$ nucleons, starting from interactions inspired by a pionless effective field theory expansion. As a benchmark, we start from model ``o'' of Ref.~\cite{Schiavilla:2021dun}, which has been shown to describe nuclear binding and radii across the nuclear chart with percent-level accuracy. We then examine the impact of charge-dependent and charge-symmetry-breaking effects, which are critical to distinguish neutron-proton, proton-proton, and neutron-neutron scattering at zero energy. We also consider the effects of $p$-wave contributions, which have been found to be important for the stability of $p$-shell nuclei~\cite{Gattobigio:2019omi}. Finally, we investigate different forms of the three-body force.

The present work is organized as follows. In Section~\ref{sec:hamiltonian}, we discuss the input Hamiltonians, including a phase-shift analysis of the baseline model ``o'', the role of charge-dependent and charge-symmetry-breaking terms, and the inclusion of $p$ waves. In Section~\ref{sec:vmc_nqs}, we describe the VMC-NQS method, providing details of the Pfaffian-Jastrow architecture, the optimization procedure, and the computational scaling of the approach. In Section~\ref{sec:results}, we discuss our results for light and medium-mass nuclei. Finally, in Section~\ref{sec:conclusions}, we summarize the main findings of our work and provide future perspectives.

\section{Input Hamiltonians}
\label{sec:hamiltonian}

\begin{figure*}[!htb]
\centering
\includegraphics[width=\textwidth]{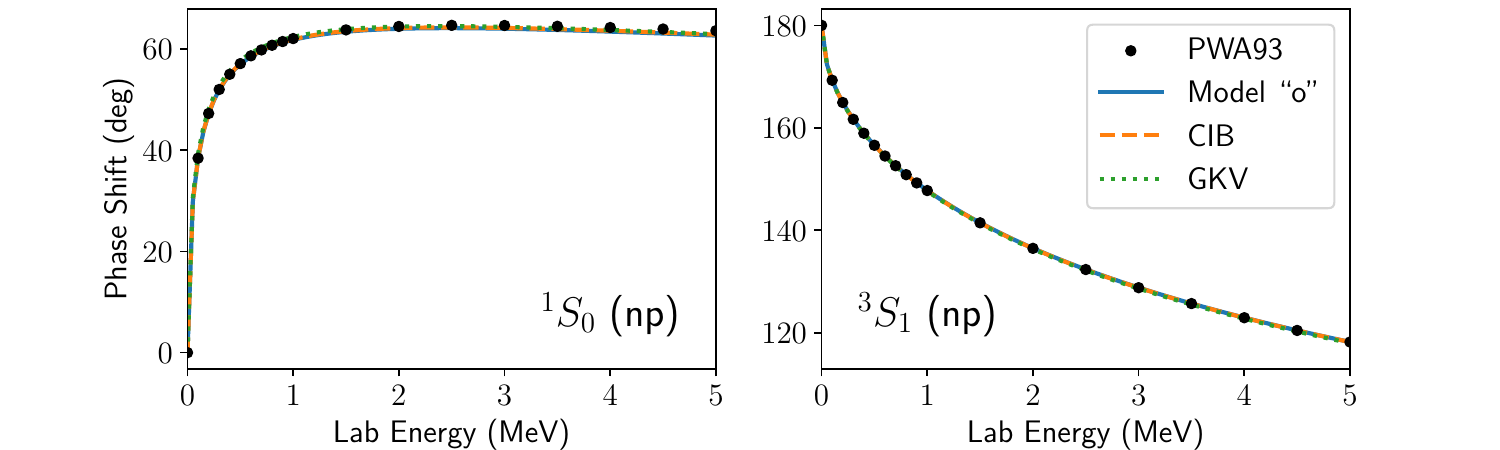}

\vspace{0.25cm}

\includegraphics[width=\textwidth]{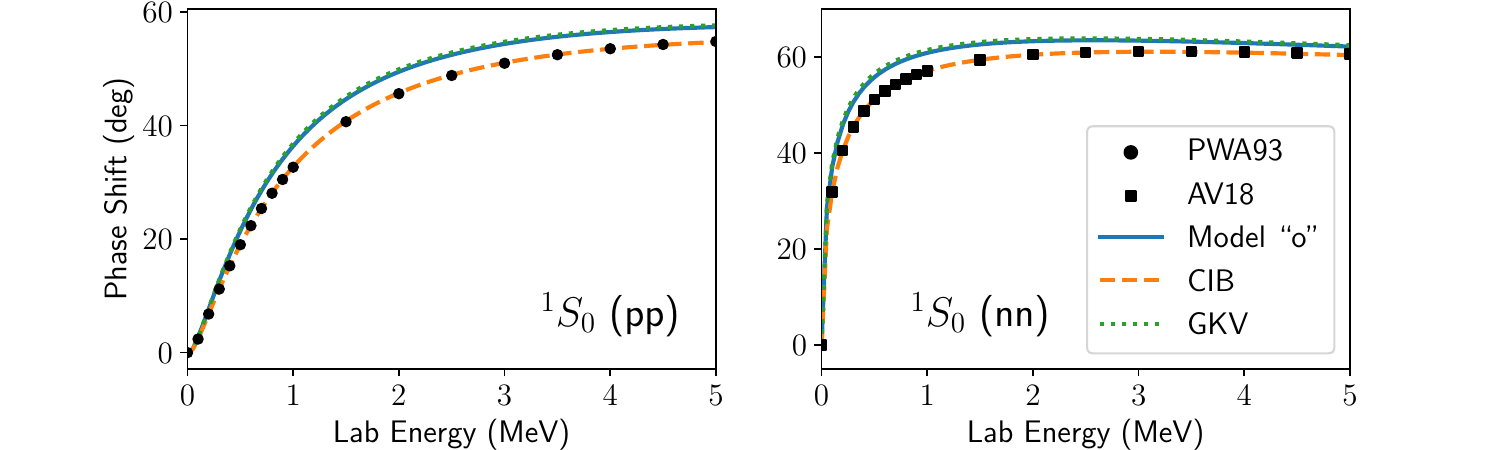}

\caption{Low-energy $s$-wave phase shifts as functions of the laboratory energy. The upper panels show neutron--proton scattering in the ${}^{1}S_{0}$ and ${}^{3}S_{1}$ channels, while the lower panels show proton--proton and neutron--neutron scattering in the ${}^{1}S_{0}$ channel. The PWA93 and AV18 results are compared with model ``o'', its CIB extension, and the GKV interaction. The GKV-weak and GKV-strong interactions give identical $s$-wave phase shifts.
}
\label{fig:s-wave-phase-shifts}
\end{figure*}

We utilize five nuclear potentials throughout this work, whose baseline is model ``o'', originally introduced in Ref.~\cite{Schiavilla:2021dun}. We consider four additional extensions, obtained by adding charge-dependent and charge-symmetry-breaking terms, incorporating $p$-wave interactions, or modifying the three-body force. 

\subsection{NN potentials}
\subsubsection{Model ``o''}
\label{sec:model-o}

Model ``o'' acts only in even partial waves, so that the coordinate-space expression of its leading-order charge-independent (CI) components is given by
\begin{equation}
v^{CI}(r_{ij}) = C_{01} v_{01}(r_{ij}) P_0^\sigma P_1^\tau                     + C_{10} v_{10}(r_{ij}) P_1^\sigma P_0^\tau ,
\label{eq:vNN_LO}
\end{equation}
where $r_{ij} = |{\bf r}_i - {\bf r}_j|$ and $P_{0,1}^\sigma$ ($P_{0,1}^\tau$) are the spin (isospin) projection operators for the nucleon pair $ij$ with total spin $S$ and isospin $T$ equal to 0 or 1:
\begin{align}
&P_0^\sigma = \frac{1 - \sigma_{ij}}{4},\qquad
P_1^\sigma = \frac{3 + \sigma_{ij}}{4},\nonumber\\
&P_0^\tau   = \frac{1 - \tau_{ij}}{4},\qquad
P_1^\tau   = \frac{3 + \tau_{ij}}{4}\, .
\end{align}
Here, $\sigma_{ij} = {\boldsymbol \sigma}_i \cdot {\boldsymbol \sigma}_j$ and $\tau_{ij} = {\boldsymbol \tau}_i \cdot {\boldsymbol \tau}_j$, with ${\boldsymbol \sigma}_i$ and ${\boldsymbol \tau}_i$ denoting the Pauli spin and isospin operators acting on nucleon $i$, respectively. In this implementation, the contact interactions are regularized using Gaussian cutoff functions,
\begin{equation}
v_{ST}(r) = \frac{1}{\pi^{3/2} R_{ST}^3} \, e^{-(r / R_{ST})^2} ,
\label{eq:gaussian-regulator}
\end{equation}
where $R_{ST}$ controls the range of the regulator. The cutoff radii $R_{01}$ and $R_{10}$, as well as the corresponding low-energy constants $C_{01}$ and $C_{10}$, were adjusted to reproduce the neutron--proton scattering lengths and effective ranges in the singlet and triplet channels, as well as the deuteron binding energy. The resulting low-energy $s$-wave phase shifts are shown in Fig.~\ref{fig:s-wave-phase-shifts}, together with those obtained from the charge-symmetry-breaking extensions discussed below.

In its original version, model ``o'' includes only one electromagnetic contribution: the Coulomb repulsion between finite-size, rather than point-like, protons.

\begin{table*}[!htb]
\caption{%
Regulator ranges and low-energy constants for the two-body potentials. The CIB-triangle interaction uses the same two-body parameters as the CIB interaction and differs only in the three-body sector. Ranges are given in fm, while the units of the coupling constants are fm$^2$.
}
\label{tab:2body-lecs}
\centering
\begin{tabular}{c|cccccccccc}
\hline\hline
Model & $R_{10}$ & $R_{01}$ & $R_{00}$ & $R_{11}$ &
$C_{10}$ & $C_{01}$ & $C_{00}$ & $C_{11}$ & $C_{CD}$ & $C_{CA}$ \\
\hline
model ``o'' & 1.546 & 1.830 & 0.0   & 0.0   & -7.040 & -5.275 & 0.0   & 0.0    & 0.0   & 0.0 \\
CIB         & 1.537 & 1.813 & 0.0   & 0.0   & -6.976 & -5.155 & 0.0   & 0.0    & 0.019 & 0.008 \\
GKV-weak    & 1.558 & 1.831 & 4.030 & 3.350 & -7.104 & -5.291 & 3.001 & -1.888 & 0.0   & 0.0 \\
GKV-strong  & 1.558 & 1.831 & 4.030 & 3.350 & -7.104 & -5.291 & 3.001 & -4.092 & 0.0   & 0.0 \\
\hline\hline
\end{tabular}
\end{table*}

\subsubsection{Charge-independence-breaking terms}
\label{sec:charge_dependent_potentials}

Relative to model ``o'', we consider an extension of the $S=0$, $T=1$ 
channel that includes charge-independence-breaking (CIB) contact terms. 
Following the terminology of the Argonne $v_{18}$ interaction, these 
consist of a charge-dependent (CD) class-II isotensor term and a 
charge-asymmetric (CA) class-III term. The CD term is written as 
$C_{CD} T_{ij}$, with
\begin{equation}
T_{ij} = 3 \tau_{iz}\tau_{jz} - \boldsymbol{\tau}_i \cdot \boldsymbol{\tau}_j ,
\end{equation}
which breaks charge independence while preserving charge symmetry. The CA term is
\begin{equation}
C_{CA}(\tau_{iz}+\tau_{jz}) ,
\end{equation}
which breaks charge symmetry and distinguishes proton--proton from 
neutron--neutron scattering. These additional contributions are restricted 
to the $T=1$ channel as they vanish in the $T=0$ channel. Hence, they can be written as
\begin{equation}
v^{\text{CIB}}(r_{ij}) = \left[ C_{CD} T_{ij} + C_{CA}(\tau_{iz}+\tau_{jz}) \right] v_{01}(r_{ij}).
\label{eq:v_LO_CIB}
\end{equation}
For compactness, we refer to the interaction obtained by adding 
$v_{\text{LO}}^{\text{CIB}}$ to $v_{\text{LO}}^{\text{CI}}$ as the CIB extension in the following.

As mentioned earlier, the only electromagnetic contribution included in model ``o''
is the one-photon Coulomb repulsion between finite-size protons. For the CIB
potentials, we additionally include the two-photon Coulomb, Darwin-Foldy,
vacuum-polarization, and spin-spin magnetic-moment terms. In the static limit, where the energy-dependent electromagnetic coupling $\alpha'$ is approximated by the fine-structure constant $\alpha$~\cite{Wiringa:1994wb}, the $pp$ terms retained are
\begin{equation}
    V_{C1}(pp) = \alpha \frac{F_C(r)}{r} ,
\end{equation}
\begin{equation}
    V_{C2} = -\frac{\alpha^2}{M_p}
    \left[ \frac{F_C(r)}{r} \right]^2 ,
\end{equation}
\begin{equation}
    V_{DF} = -\frac{\alpha}{4M_p^2} F_\delta(r) ,
\end{equation}
\begin{equation}
    V_{VP} =
    \frac{2\alpha^2}{3\pi}\frac{F_C(r)}{r} I_{VP}(r) ,
\end{equation}
and
\begin{equation}
    V_{MM}(pp) =
    -\frac{\alpha \mu_p^2}{6 M_p^2}
    F_\delta(r) \boldsymbol{\sigma}_i\cdot\boldsymbol{\sigma}_j .
\end{equation}

For $np$ interactions, we include the Coulomb term associated with the neutron
charge distribution and the spin-spin magnetic-moment term,
\begin{equation}
    V_{C1}(np) = \alpha \beta_n \frac{F_{np}(r)}{r} ,
\end{equation}
\begin{equation}
    V_{MM}(np) =
    -\frac{\alpha \mu_n\mu_p}{6 M_n M_p}
    F_\delta(r) \boldsymbol{\sigma}_i\cdot\boldsymbol{\sigma}_j .
\end{equation}
Finally, for $nn$ interactions, we include the spin-spin magnetic-moment term,
\begin{equation}
    V_{MM}(nn) =
    -\frac{\alpha \mu_n^2}{6 M_n^2}
    F_\delta(r) \boldsymbol{\sigma}_i\cdot\boldsymbol{\sigma}_j .
\end{equation}

\begin{figure*}[!htb]
\centering
\includegraphics[width=\textwidth]{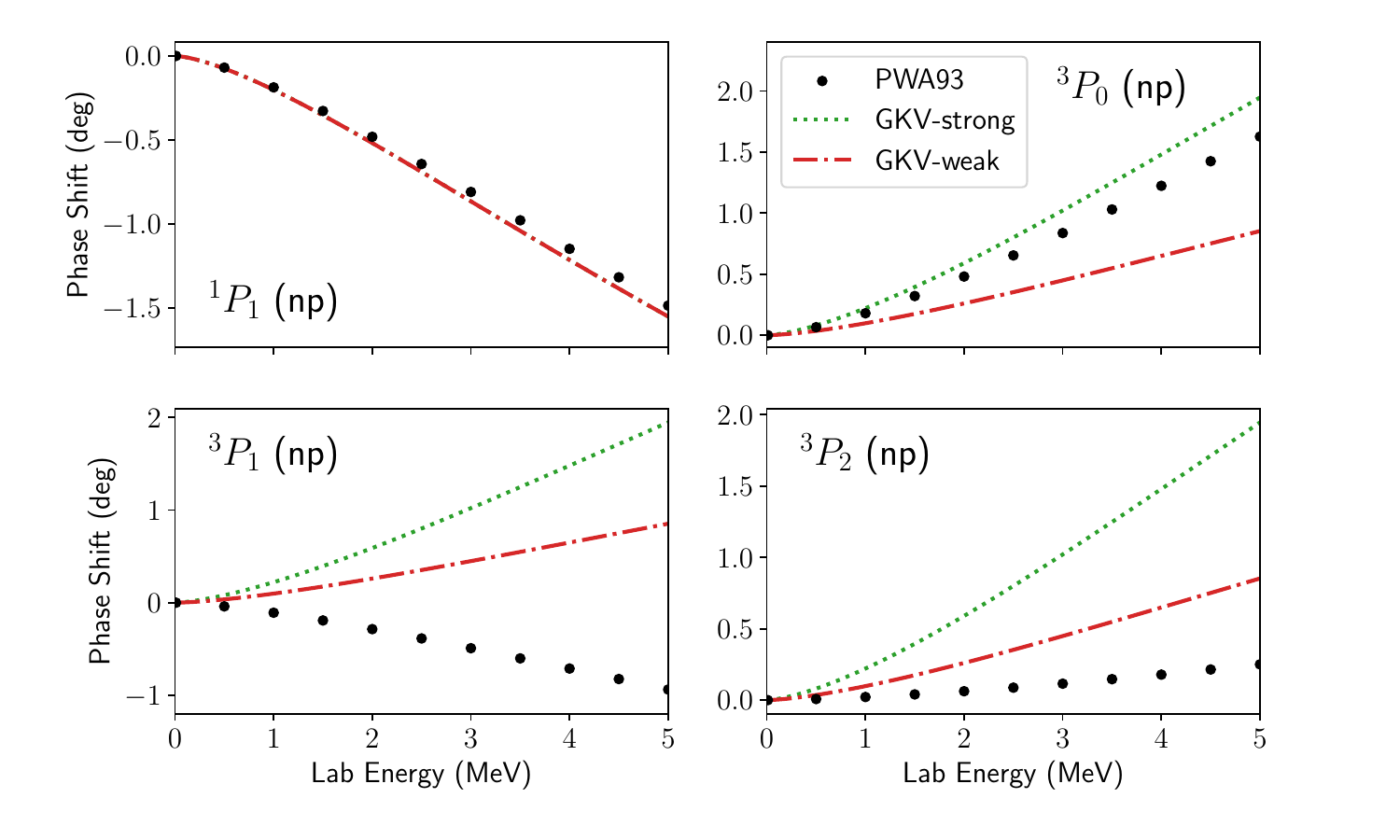}
\caption{Phase shifts for neutron--proton scattering in the ${}^1P_1$ and ${}^3P_J$ $(J=0,1,2)$ channels as functions of the laboratory energy $T_{\mathrm{lab}}$. PWA93 phase shifts are compared with the GKV interaction for the strong and weak choices of $V_{11}$. }
\label{fig:np-p-wave}
\end{figure*}

The finite-size functions are taken from Ref.~\cite{Wiringa:1994wb}. With
$b=4.27~\mathrm{fm}^{-1}$ and $x \equiv br$, they read
\begin{equation}
    F_C(r) =
    1-\left(1+\frac{11}{16}x+\frac{3}{16}x^2+\frac{1}{48}x^3\right)e^{-x} ,
\end{equation}
\begin{equation}
    F_{np}(r) =
    b^2 \left(15x+15x^2+6x^3+x^4\right)\frac{e^{-x}}{384} ,
\end{equation}
and
\begin{equation}
    F_\delta(r) =
    b^3\left(\frac{1}{16}+\frac{1}{16}x+\frac{1}{48}x^2\right)e^{-x} .
\end{equation}
The vacuum-polarization function is
\begin{equation}
     I_{VP}(r) =
    \int_1^\infty dx\, e^{-2m_e r x/(\hbar c)}
    \left(1+\frac{1}{2x^2}\right)
    \frac{(x^2-1)^{1/2}}{x^2} .
\end{equation}
In practice, we evaluate it using the short-range approximation
\begin{equation}
    I_{VP}(r) \simeq
    -\gamma - \frac{5}{6}
    + \left|\log\left(\frac{m_e r}{\hbar c}\right)\right|
    + \frac{3\pi}{4}\frac{m_e r}{\hbar c}.
\end{equation}

Here, $M_p$ and $M_n$ are the proton and neutron masses, $\mu_p$ and $\mu_n$ are the proton and neutron magnetic moments, $m_e$ is the electron mass, and $\beta_n$ is the slope of the neutron electric form factor at zero momentum transfer. 
Relative to Ref.~\cite{Wiringa:1994wb}, we retain only the spin-spin part of the magnetic moment interaction. The tensor and spin-orbit magnetic-moment terms do not
contribute to the $pp$ ${}^1S_0$ channel considered here. We also adopt the approximation $\alpha'=\alpha$, consistent with the static electromagnetic interaction used in model ``o''.

The resulting two-nucleon interaction is fit to reproduce low-energy phase shifts in the $S=0$, $T=1$ and $S=1$, $T=0$ channels. We use phase shifts from the Nijmegen partial-wave analysis (PWA93)~\cite{Stoks:1993tb} for $np$ and $pp$ scattering, and phase shifts computed from the Argonne $v_{18}$ potential~\cite{Wiringa:1994wb} for $nn$ scattering, all up to laboratory energies of $5$ MeV. The fits are performed independently in each spin-isospin channel. For the neutral $np$ and $nn$ channels, the phase shifts are extracted by matching the radial wave functions to free spherical Bessel functions. For $pp$ scattering, where the one-photon Coulomb interaction is long ranged, we instead match to regular and irregular Coulomb wave functions, thereby extracting the nuclear phase shift relative to the Coulomb asymptotic state.

As shown in the lower two panels of Fig.~\ref{fig:s-wave-phase-shifts}, the inclusion of CIB terms accounts for the splitting between the proton--proton and neutron--neutron phase shifts at low energies, which are degenerate in model ``o''. For neutron--neutron scattering, where no PWA93 phase shifts are available, AV18 is used as the reference. The resulting LECs and regulator cutoffs are listed in Table~\ref{tab:2body-lecs}.

\subsubsection{Gattobigio-Kievsky-Viviani potential}
In Ref.~\cite{Gattobigio:2019omi}, it was found that, for the $s$-wave interaction considered there, $^6$He and $^6$Li were unbound once the repulsive three-body force and Coulomb interaction were included. They further showed that adding a weak attractive $p$-wave component was sufficient to bind the six-nucleon systems. 
Following that work, we extend the CI model ``o'' interaction of Eq.~\eqref{eq:vNN_LO} by including the additional spin-isospin channels that contribute to odd partial waves,
\begin{align}
v^{CI}(r_{ij}) &= C_{01} v_{01}(r_{ij}) P_0^\sigma P_1^\tau
+ C_{10} v_{10}(r_{ij}) P_1^\sigma P_0^\tau \nonumber\\
&+ C_{11} v_{11}(r_{ij}) P_1^\sigma P_1^\tau
+ C_{00} v_{00}(r_{ij}) P_0^\sigma P_0^\tau\,.
\label{eq:vNN_p-waves}
\end{align}
The couplings and ranges of this interaction are also listed in Table~\ref{tab:2body-lecs}. The original GKV interaction is written in terms of unnormalized Gaussian strengths, $V_{ST}\exp[-(r/r_{ST})^2]$, whereas Table~\ref{tab:2body-lecs} reports
the corresponding couplings in the normalization convention of Eq.~\eqref{eq:gaussian-regulator}. As in Ref.~\cite{Gattobigio:2019omi}, the $S=0$, $T=0$ channel is fixed
to reproduce the low-energy ${}^1P_1$ phase shift, shown in the upper-left panel of Fig.~\ref{fig:np-p-wave}.

Since the interaction of Eq.~\eqref{eq:vNN_p-waves} contains no tensor or spin-orbit operators, it cannot distinguish the different total-angular-momentum states in the two-body triplet $p$ waves. Therefore, in the $S=1$, $T=1$ channel we use a Gaussian interaction with range $r_{11}=3.35$ fm and consider two representative strengths,
denoted GKV-weak and GKV-strong. These correspond to the original GKV choices $V_{11}=-1.78$ MeV and $V_{11}=-3.857$ MeV, respectively; their values in the convention of Eq.~\eqref{eq:gaussian-regulator} are given in Table~\ref{tab:2body-lecs}. As shown in the remaining panels of Fig.~\ref{fig:np-p-wave}, these two choices generate the same central-interaction trend in the ${}^3P_J$ channels and therefore
cannot reproduce the full $J$ dependence of the AV18 and PWA93 phase shifts.

\subsection{Three-body forces}
\label{sec:three-body}

\subsubsection{Model ``o''}
The three-nucleon ($3N$) force included in model ``o'' is of the form
\begin{align}
    V_{ijk} = \frac{c_E}{f_\pi^4 \Lambda_\chi} \frac{(\hbar c)^6}{\pi^3 R_3^6} 
    \sum_{\rm cyc} e^{-(r_{ij}^2 + r_{jk}^2)/R_3^2} ,
    \label{eq:o_3NF}
\end{align}
where $\Lambda_\chi = 1$ GeV is the breaking scale, $f_\pi = 92.4$ MeV is the pion decay constant, and $\sum_{\rm cyc}$ denotes cyclic permutations of the indices $i$, $j$, and $k$. The low-energy constant $c_E$ is fixed to reproduce the $^3$H binding energy for a given value of the cutoff $R_3$. The analysis of Ref.~\cite{Schiavilla:2021dun} shows that the choice $R_3 = 1.0$ fm and $c_E = 1.0786$ provides a satisfactory description of nuclear binding energies across a broad mass range, up to $^{90}$Zr. However, subsequent VMC-NQS calculations have indicated that this choice leads to overbinding in $^{16}$O and heavier nuclei. Increasing the cutoff to $R_3 = 1.1$ fm with $c_E = 1.2945$ largely resolves this issue~\cite{Gnech:2023prs}, as the extended range of the $3N$ force introduces additional repulsion in heavier systems. In this work, when referring to model ``o'', we always mean the version with $R_3 = 1.1$ fm, unless otherwise stated.

\subsubsection{CIB and GKV potentials}
The three-body force associated with the CIB and GKV $NN$ potentials retains the same functional form as in Eq.~\eqref{eq:o_3NF}. However, as argued in Ref.~\cite{Lynn:2017fxg}, regulator artifacts in contact $3N$ forces are reduced by retaining only triples with total spin and isospin $S=1/2$, $T=1/2$, which survive the zero-range limit. A similar form was also considered in Ref.~\cite{Weiss-Attia:2024tee} in the context of perturbative applications of next-to-leading-order pionless EFT in a finite volume. Employing the full spin-isospin projector is computationally expensive, however, because it is not diagonal in spin-isospin space. For a generic triple, it couples a configuration to as many as three spin states and three isospin states, requiring up to nine wave-function amplitudes, or eight additional wave-function evaluations if the original amplitude is already available.

As a computationally efficient alternative, we multiply each triple contributing to the $3N$ force of Eq.~\eqref{eq:o_3NF} by the simplified diagonal filter
\begin{align}
    P_{ijk} &=
    \frac{1}{16}
    (3-s_{zi} s_{zj} - s_{zi} s_{zk} - s_{zj} s_{zk})\nonumber\\
    &\times (3-t_{zi} t_{zj} - t_{zi} t_{zk} - t_{zj} t_{zk}) ,
\label{eq:three_body_proj}
\end{align}
where $s_{zi}=\pm 1$ and $t_{zi}=\pm 1$ denote the spin and isospin projections of particle $i$ along the $z$-axis. With this choice, triples in which all three particles have the same spin projection or the same isospin projection, such as $nnn$, $ppp$, $\uparrow\uparrow\uparrow$, and $\downarrow\downarrow\downarrow$, are removed from the $3N$ interaction.

For the CIB interaction only, we also consider a structural modification of the three-body force. In Eq.~\eqref{eq:o_3NF}, the three-body interaction is written as a sum of products of two pairwise Gaussian factors. As a result, a given term can remain large when two pair distances are short, even if the third pair distance is relatively large. This can occur, for example, in a chain-like arrangement where particles $i$ and $j$ are close and particles $j$ and $k$ are close, while particles $i$ and $k$ are farther apart. As an alternative, we consider the hypercentral three-body potential of Ref.~\cite{Gattobigio:2019omi},
\begin{equation}
    V_{ijk} = W_0 \exp\left[-\frac{r_{ij}^2 + r_{ik}^2 + r_{jk}^2}{R_3^2}\right]\,,
\end{equation}
which contributes appreciably only when all three pair distances are short. This form therefore favors compact triangular arrangements of the three particles and suppresses chain-like ones. We expect this geometry to reduce the repulsion generated by triples involving two halo nucleons and one core nucleon in nuclei such as $^6$Li and $^6$He. We denote this interaction as ``CIB-triangle''. The same diagonal filter $P_{ijk}$ is applied in this case.

The strength and range parameters of the $3N$ interactions are determined through an iterative search. For each trial set of parameters, we compute NQS ground-state energies for $^4$He and $^{16}$O using the full Hamiltonian, and adjust the three-body parameters until both energies fall within the NQS uncertainties. The resulting three-body parameters, listed in Table~\ref{tab:3body}, are then held fixed in the calculations of all other nuclei.

\begin{table}[htb]
\caption{Strengths and ranges of the $3N$ force. The coefficient $c_E$ is dimensionless. For the CIB-triangle interaction, $W_0$ denotes the effective coefficient of the hypercentral Gaussian and is given in MeV.}
\label{tab:3body}
\centering
\begin{tabular}{lccc}
    \hline\hline
    Name         & $c_E$   & $W_0$ (MeV) & $R_3$ (fm) \\
    \hline
    Model ``o''  & 1.295   & ---    & 1.100 \\
    CIB          & 1.870   & ---    & 1.322 \\
    CIB-triangle & ---     & 22.36  & 1.793 \\
    GKV-weak     & 4.487   & ---    & 1.734 \\
    GKV-strong   & 10.538  & ---    & 2.175 \\
    \hline\hline
\end{tabular}
\end{table}

As shown in Table~\ref{tab:3body}, the CIB interaction requires both a larger strength and a larger range for the $3N$ force than model ``o''. This behavior is primarily due to the diagonal filter $P_{ijk}$ of Eq.~\eqref{eq:three_body_proj}. For the spin-isospin configurations relevant to $^4$He, this filter leaves all triples active and therefore does not modify the $3N$ contribution. In $^{16}$O, however, the filter removes triples in which all three particles have the same spin projection or the same isospin projection. For fixed values of the $3N$ parameters, this suppresses the repulsive $3N$ matrix element; the fit therefore compensates by increasing both $c_E$ and $R_3$. The GKV interactions require even larger equivalent $c_E$ values. This reflects the additional attraction generated by the $p$-wave terms, which increases the binding of $^{16}$O more strongly than that of $^4$He. A more repulsive and longer-ranged $3N$ force is therefore needed to reproduce both calibration nuclei, with the largest effect occurring for the GKV-strong parametrization.

\section{Neural-network quantum state architecture}
\label{sec:vmc_nqs}
Following early applications to spin systems~\cite{Carleo:2016svm} and quantum chemistry~\cite{Hermann:2022}, neural-network quantum states (NQS) have proven highly effective for solving the nuclear Schr\"odinger equation, with applications ranging from the deuteron to $^{20}$Ne~\cite{Keeble:2019bkv,Adams:2020aax,Gnech:2021wfn,Lovato:2022tjh,Yang:2022rlw,Gnech:2023prs,Yang:2025mhg}, as well as to neutron matter~\cite{Fore:2022ljl}.
More recently, a compact and flexible neural Pfaffian-Jastrow (PJ) architecture has been introduced, in which a message-passing neural network (MPNN) efficiently encodes pairing and backflow correlations. This approach has been shown to match or surpass state-of-the-art diffusion Monte Carlo calculations for ultracold Fermi gases~\cite{Kim:2023fwy}, and to be particularly effective at capturing pairing effects relevant to low-density nucleonic matter, such as that found in neutron-star crusts~\cite{Fore:2024exa}. It has also recently been applied to hypernuclei~\cite{DiDonna:2025oqf}.

\subsection{Pfaffian Jastrow Ansatz}
The PJ+MPNN ansatz is schematically expressed as in Ref.~\cite{Kim:2023fwy},
\begin{equation}
\Psi_{\text{PJ}}(X) = e^{J(X)} \mathrm{pf}(\Phi(X)) ,
\label{eq:PJ}
\end{equation}
where we denote the set of single-particle variables by
\begin{equation}
X = (\mathbf{x}_1,\dots,\mathbf{x}_A).
\end{equation}
Here, each $\mathbf{x}_i=(\mathbf{r}_i,\mathbf{s}_i)$ comprises the spatial Cartesian coordinates $\mathbf{r}_i$ and the spin--isospin projections $\mathbf{s}_i=(s_{z,i},t_{z,i})$ of the $i$th particle. We use the convention $t_z=1$ for protons and $t_z=-1$ for neutrons. Following Ref.~\cite{Massella:2018xdj}, we enforce translational invariance by replacing
\begin{equation}
\mathbf{r}_i \to \mathbf{r}_i-\mathbf{R}_{\text{CM}},
\qquad
\mathbf{R}_{\text{CM}}=\frac{1}{A}\sum_{i=1}^A \mathbf{r}_i ,
\end{equation}
where $\mathbf{R}_{\text{CM}}$ denotes the center-of-mass coordinate.

The complex-valued, permutation-invariant Jastrow factor is defined as
\begin{equation}
J(X) = a\,\tanh\left(\frac{U_J(X)}{a}\right) + i\,V_J(X)\,.
\label{eq:reg_jas}
\end{equation}
Here, \(U_J(X)\) and \(V_J(X)\) are real-valued functions representing the logarithmic amplitude and phase, respectively. The parameter \(a\) acts as a cutoff that regularizes the growth of \(U_J\) and helps mitigate potential runaway instabilities. Following Ref.~\cite{Bukov:2021mnn}, we choose \(a = 8\), corresponding to a maximum relative magnitude variation of approximately \(10^{7}\). We enforce both $U_J(X)$ and $V_J(X)$ to be permutation invariant by employing the Deep Sets architecture~\cite{Zaheer:2017,Wagstaff:2019} with {\it logsumexp} pooling.

The antisymmetric part of the nucleonic wave function is given by the Pfaffian of a skew-symmetric matrix. For nuclei with an even number of nucleons, it is given by
\begin{equation}
   \Phi(X)= \begin{bmatrix}
0 & \phi(\mathbf{x}_1, \mathbf{x}_2) & \dots & \phi(\mathbf{x}_1, \mathbf{x}_{A}) \\
\phi(\mathbf{x}_2, \mathbf{x}_1) & 0 & \dots & \phi(\mathbf{x}_2, \mathbf{x}_{A}) \\
\vdots & \vdots & \ddots & \vdots \\
\phi(\mathbf{x}_{A}, \mathbf{x}_1) & \phi(\mathbf{x}_{A}, \mathbf{x}_2) & \dots & 0
\end{bmatrix}\,.
    \label{eq:wavefunction_pfaffian}
\end{equation}
To ensure skew-symmetry, the pairing orbital is defined as
\begin{equation}
    \phi(\mathbf{x}_i, \mathbf{x}_j)= 
    \eta(\mathbf{x}_i, \mathbf{x}_j) - \eta(\mathbf{x}_j, \mathbf{x}_i),
    \label{eq:skew_symmetric_elements}
\end{equation}
where $\eta(\mathbf{x}_i, \mathbf{x}_j)$ is a complex-valued function, whose logarithm is regularized as in Eq.~\eqref{eq:reg_jas} 
\begin{equation}
    \eta(\mathbf{x}_i, \mathbf{x}_j)=\exp\left[a\,\tanh\left(\frac{u_\eta(\mathbf{x}_i, \mathbf{x}_j)}{a}\right)+i \pi\,v_\eta(\mathbf{x}_i, \mathbf{x}_j)\right]\, .
\end{equation}
The real-valued multilayer perceptrons (MLPs) $u_\eta$ and $v_\eta$  encode the logarithmic amplitude and phase of $\eta$.

Nuclei with an odd number of nucleons are treated by adding to the Pfaffian an unpaired single-particle orbital~\cite{Kim:2023fwy}. Thus, we extend $\Phi(X)$ to an $(A+1)\times (A+1)$ skew-symmetric matrix $\tilde{\Phi}(X)$ by introducing an additional row and column:
\begin{equation}
    \tilde{\Phi}= \begin{bmatrix}
 \Phi(X_A) & \mathbf{u} \\
-\mathbf{u}^T & 0
\end{bmatrix}\,,
    \label{eq:wavefunction_pfaffian_odd}
\end{equation}
where
\begin{equation}
     \mathbf{u}=\begin{bmatrix}
 \psi(\mathbf{x}_1) \\
\psi(\mathbf{x}_2)\\
\vdots \\
\psi(\mathbf{x}_A)
\end{bmatrix}\,.
\end{equation}
Consistent with the complex-valued functions above, we use two separate real-valued neural networks to parameterize the logarithm of $\psi(\mathbf{x}_i)$:
\begin{align}
\psi(\mathbf{x})=\exp\left[a\,\tanh\left(\frac{u_\psi(\mathbf{x})}{a}\right)+i \pi\,v_\psi(\mathbf{x})\right]\,.
    \label{eq:psi_single}
\end{align}
We again set \(a = 8\) to mitigate potential runaway instabilities and both $u_\psi$ and $v_\psi$ are MLPs.

\subsection{Backflow transformation}
We employ an MPNN backflow transformation that takes as input single-particle and pairwise visible features
\begin{equation}
\mathbf{v}_i = (\mathbf{r}_i-\mathbf{R}_{\text{CM}},\mathbf{s}_i),\quad \mathbf{v}_{ij} = [\mathbf{r}_{ij}, r_{ij}, \mathbf{s}_i,\mathbf{s}_j], 
\end{equation}
The initial hidden node and edge features are obtained by concatenating the visible features with learned embeddings,
\begin{equation}
\mathbf{h}_i^0 = [\mathbf{v}_i, f_A(\mathbf{v}_i)] ,
\qquad
\mathbf{h}_{ij}^0 = [\mathbf{v}_{ij}, f_B(\mathbf{v}_{ij})] .
\end{equation}
As in Refs.~\cite{Fore:2024exa,DiDonna:2025oqf}, the functions $f_A$ and $f_B$ are implemented as MLPs, rather than as the linear layers employed in the unitary Fermi-gas work of Ref.~\cite{Kim:2023fwy}. These functions ensure that the dimensions of the hidden features $\mathbf{h}_i^t$ and $\mathbf{h}_{ij}^t$ remain fixed across message-passing layers.

The MPNN update is performed iteratively for $t=1,\dots,T$. At each layer, information is exchanged between node and edge features through messages
\begin{equation}
\mathbf{m}_{ij}^t = f_M^t(\mathbf{h}_i^{t-1},\mathbf{h}_{ij}^{t-1},\mathbf{h}_j^{t-1}) .
\end{equation}
For each particle $i$, the messages are collected and pooled over all $j\neq i$, removing any dependence on the ordering of the other particles. As in Ref.~\cite{Kim:2023fwy}, we use logsumexp pooling, a smooth alternative to max pooling:
\begin{equation}
\mathbf{m}_{i}^t = \log\left(\sum_{j\neq i}\exp\left(\mathbf{m}_{ij}^t\right)\right),
\end{equation}
where the logarithm and exponential are applied componentwise. The hidden node and edge features are then updated as
\begin{align}
\mathbf{h}_i^t &= [\mathbf{v}_i, f_F^t(\mathbf{h}_i^{t-1},\mathbf{m}_{i}^t)] , \\
\mathbf{h}_{ij}^t &= [\mathbf{v}_{ij}, f_G^t(\mathbf{h}_{ij}^{t-1},\mathbf{m}_{ij}^t)] .
\end{align}
The functions $f_M^t$, $f_F^t$, and $f_G^t$ are distinct fully connected neural networks, with output dimensions matching those of $f_A$ and $f_B$. The concatenated skip connections to the visible features ensure that the raw input remains accessible as the MPNN depth $T$ increases.

After the final message-passing layer, the hidden node and edge features are aggregated into pairwise and single-particle feature vectors,
\begin{align}
\mathbf{g}_{ij} &= [\mathbf{h}_i^T,\mathbf{h}_j^T,\mathbf{h}_{ij}^T] , \\
\mathbf{g}_i &= \log\left(\sum_{j\neq i}\exp\left(\mathbf{g}_{ij}\right)\right).
\end{align}
By construction, the collection $\{\mathbf{g}_i\}$ is permutation equivariant: under a permutation of particle labels, $\mathbf{g}_i$ transforms with the label $i$. Equivalently, for fixed $i$, $\mathbf{g}_i$ is invariant under exchanges $\mathbf{x}_j\leftrightarrow\mathbf{x}_k$ for all $j,k\neq i$. Similarly, $\mathbf{g}_{ij}$ is invariant under exchanges $\mathbf{x}_l\leftrightarrow\mathbf{x}_m$ for all $l\neq i,j$ and $m\neq i,j$. We use a single message-passing layer, $T=1$. Consistent with related continuous-space studies~\cite{Kim:2023fwy}, $T=1$ already captures most of the energy gain, while $T=2$ can yield small, system-dependent improvements.

The pairwise feature vector $\mathbf{g}_{ij}$ is provided as input to both the Pfaffian and the Jastrow factor, replacing the original pairwise inputs $(\mathbf{x}_i,\mathbf{x}_j)$ with learned backflow features. Likewise, for odd-$A$ calculations, the single-particle feature vectors $\mathbf{g}_i$ are used in place of the original inputs $\mathbf{x}_i$ for the unpaired orbital in Eq.~\eqref{eq:wavefunction_pfaffian_odd}. All MLPs in the ansatz have two hidden layers, each with width 16. The output dimension is either 1 or 16, depending on whether the network maps to the final output space or to a latent space. The activation function used throughout is the hyperbolic tangent, and all pooling operations use logsumexp. In total, the neural-network quantum state contains $11,012$ trainable parameters for the even-$A$ systems and 13,734 for the odd-$A$ systems.

\subsection{Wave function optimization}
The optimal parameters of the NQS are obtained by minimizing the variational energy,
\begin{equation}
E_V(\mathbf{p}) = \langle H \rangle_{\mathbf{p}} = \frac{\langle \Psi_V(\mathbf{p}) | H | \Psi_V(\mathbf{p}) \rangle}{\langle \Psi_V(\mathbf{p}) | \Psi_V(\mathbf{p}) \rangle} .
\label{eq:variational-energy}
\end{equation}
The minimization is performed using the stochastic reconfiguration (SR) algorithm~\cite{Sorella:2005}. We define the derivative operator $O_i$ through its action on the variational state,
\begin{equation}
O_i |\Psi_V(\mathbf{p})\rangle = \frac{\partial}{\partial p_i} |\Psi_V(\mathbf{p})\rangle .
\label{eq:Oi-definition}
\end{equation}
For real variational parameters, the energy gradient can be written as
\begin{equation}
g_i \equiv \frac{\partial E_V}{\partial p_i} = 2 \operatorname{Re} \Big[ \langle O_i^\dagger H \rangle_{\mathbf{p}} - E_V(\mathbf{p})\langle O_i^\dagger \rangle_{\mathbf{p}} \Big] .
\label{eq:gradient}
\end{equation}
The SR matrix is the covariance matrix of the wave-function derivatives,
\begin{equation}
S_{ij} = \langle O_i^\dagger O_j \rangle_{\mathbf{p}} - \langle O_i^\dagger \rangle_{\mathbf{p}} \langle O_j \rangle_{\mathbf{p}} .
\label{eq:sr-matrix}
\end{equation}
Up to conventional factors, $S$ corresponds to the quantum Fisher information matrix~\cite{Bukov:2021mnn}.

At optimization step $t$, we compute the gradient $\mathbf{g}_t$ and SR matrix $S_t$. For complex wave functions defined in terms of real parameters, a suitable update can be obtained by solving
\begin{equation}
\Delta\mathbf{p}_t = -\eta (\operatorname{Re} S_t)^{-1} \mathbf{g}_t ,
\label{eq:sr-update}
\end{equation}
where $\eta$ is the learning rate. To stabilize the inversion, we use an RMSProp-style running average of the squared gradients~\cite{Lovato:2022tjh},
\begin{equation}
\mathbf{v}_t = \beta \mathbf{v}_{t-1} + (1-\beta) \mathbf{g}_t^2 .
\end{equation}
We take $\beta=0.99$, with the square taken componentwise, and replace
\begin{equation}
S_t \to S_t + \varepsilon \operatorname{diag}\Big(\sqrt{\mathbf{v}_t}+10^{-8}\Big) ,
\end{equation}
where $\varepsilon>0$ is a small constant, typically set to $10^{-3}$.

Convergence to the nuclear ground state is typically achieved in $\mathcal{O}(10^3)$ optimization steps, with the most demanding case, $^{58}\mathrm{Ni}$, requiring 4800 steps. To reduce the computational cost for nuclei with $A \geq 28$, we employ transfer learning across different nuclear potentials. Specifically, we first obtain a converged ground state for a given potential and then use the trained model as the initialization for other potentials of the same nucleus, which require significantly fewer optimization steps to converge~\cite{Kim:2023fwy,DiDonna:2025oqf}. In most cases, retraining requires approximately 300 additional steps.

We validate this procedure in $^{40}\mathrm{Ca}$ by training separate models for the charge-dependent potentials without transfer learning and comparing the results with those obtained using transfer learning. The resulting total binding energies agree within about 1 MeV, corresponding to a difference at the 0.4\% level, providing confidence in the accuracy of the transfer-learning approach.

\begin{figure}[!t ]
    \includegraphics[width=0.95\linewidth]{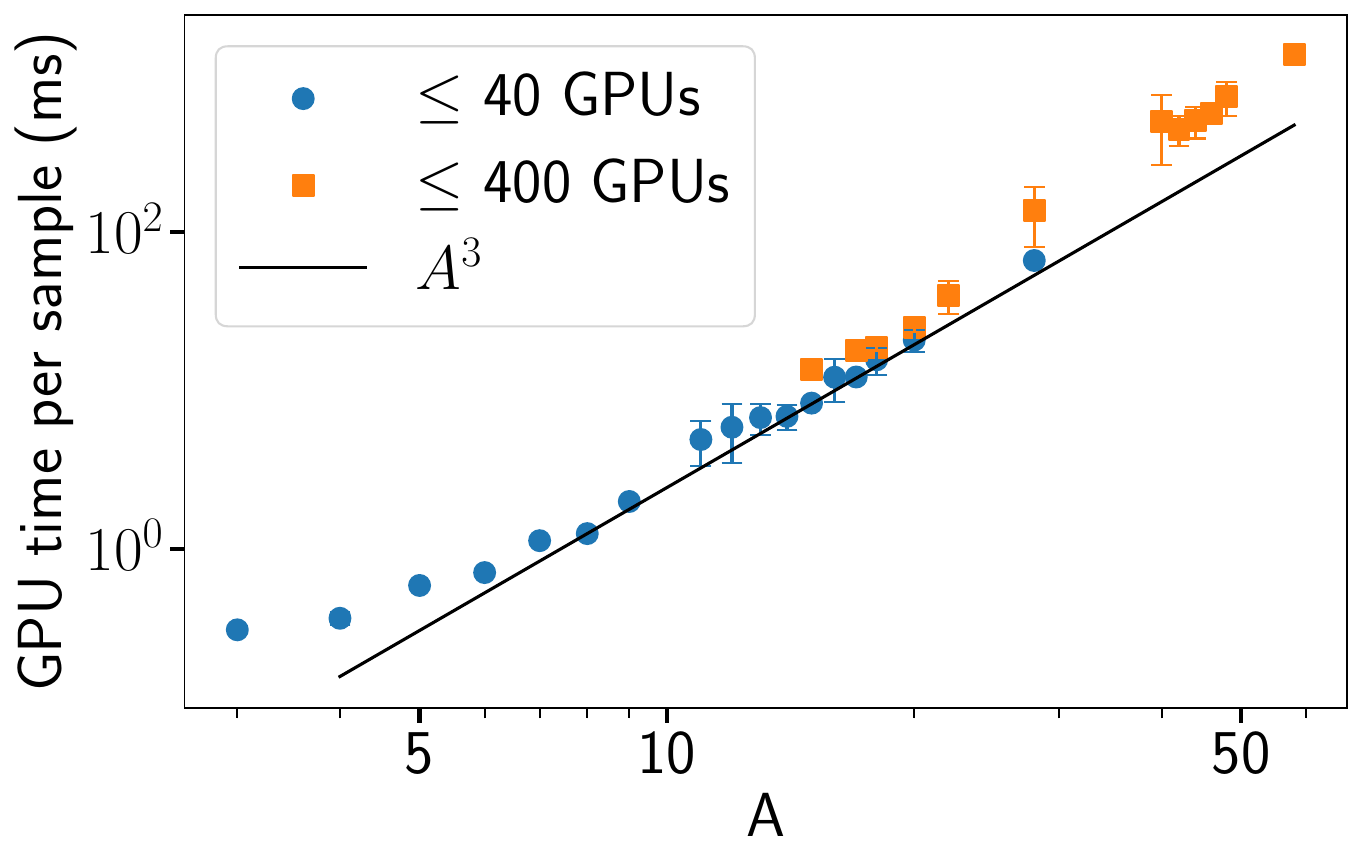}
    \caption{Scaling of the NQS algorithm with increasing size of nucleus. The runtime for each point is scaled proportionally to the number of GPUs used and inversely to the number of samples. Different markers are used to distinguish ranges of GPU numbers. The black line indicates the A$^3$ behavior.}
    \label{fig:time_vs_A}
\end{figure}

\begin{figure}[!b]
    \includegraphics[width=0.95\linewidth]{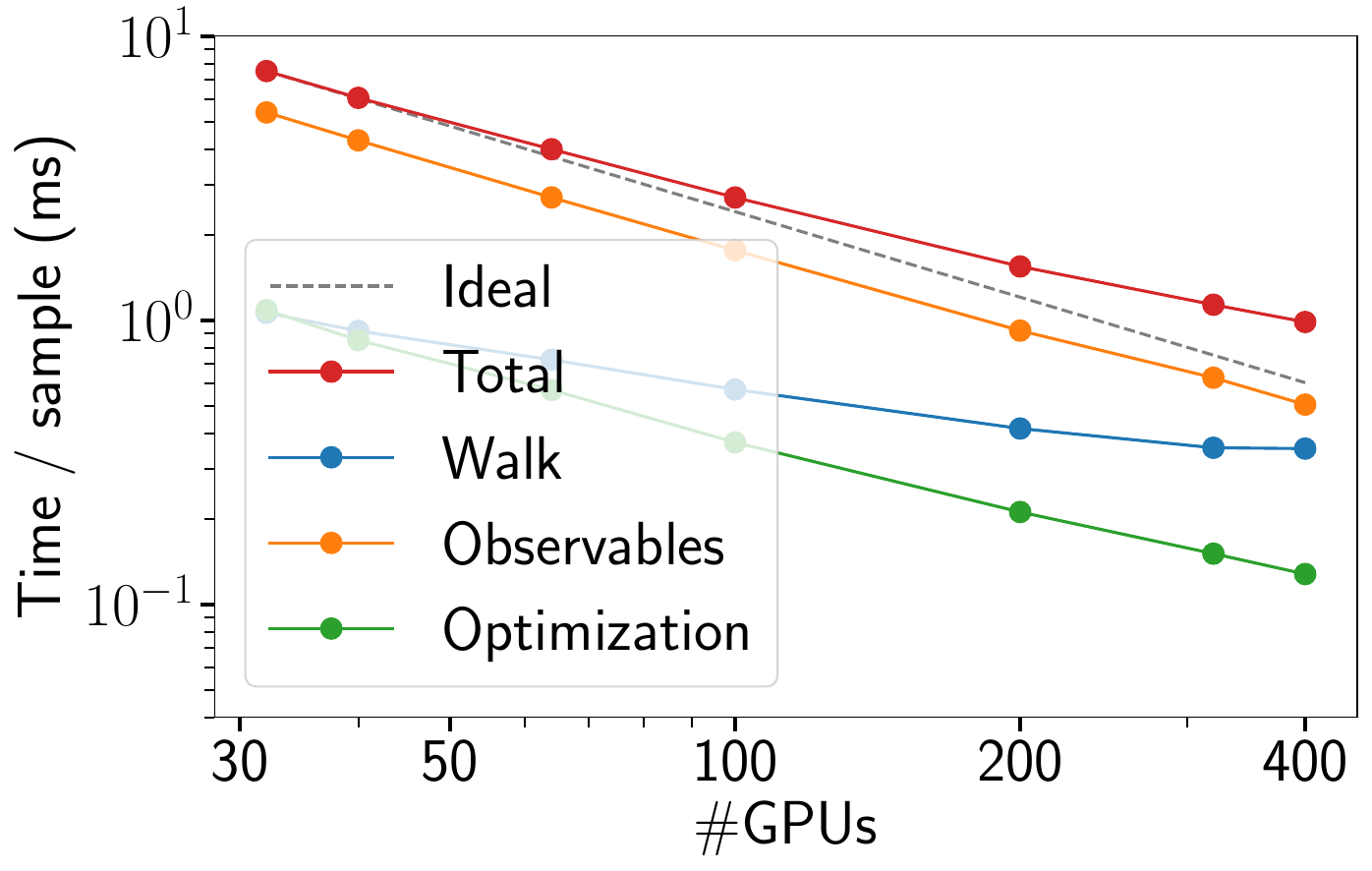}
    \caption{NQS scaling with number of GPUs used as measured on a calculation of $^{40}$Ca. The time per sample is computed by dividing the average time for single step of the training algorithm by the number of samples used in that training step. All data points are computed with $16000$ samples.}
    \label{fig:log_scaling}
\end{figure}

\subsection{Computational Scaling}

NQS offer an efficient representation of the physically relevant region of Hilbert space, leading to polynomial computational scaling with system size. In the calculations presented here, the evaluation of the Pfaffian wave function scales as $A^3$, while the computation of spin--isospin-dependent potentials, which involves repeated wave-function evaluations, scales as $A^5$~\cite{Lovato:2022tjh}.

All simulations are performed in parallel on NVIDIA A100 GPUs using a minimum of 16{,}000 Monte Carlo samples distributed across 1600 independent chains. The wall-clock cost per optimization step increases with the size of the nucleus and with the number of GPUs required, as larger systems demand more resources due to memory and time constraints.

\begin{table*}[!htb]
\centering
\caption{Average deviations from experiment for the ground-state energy per particle, $\Delta E/A$, and charge radius, $\Delta r_{\mathrm{ch}}$ for the different interactions employed in this work.}
\label{tab:avg_deviation}
\setlength{\tabcolsep}{14pt} 
\begin{tabular}{l c c}
\hline
\hline
Potential & $\Delta E/A$ [MeV] & $\Delta r_{\mathrm{ch}}$ [fm] \\
\midrule
Model ``o''   & 0.306(19) & 0.127(97) \\
CIB           & 0.322(7)  & 0.149(5)  \\
CIB-triangle  & 0.306(24) & 0.146(5)  \\
GKV-weak      & 0.223(6)  & 0.138(5)  \\
GKV-strong    & 0.257(10) & 0.332(11) \\
\hline
\hline
\end{tabular}
\end{table*}

To assess the overall scaling behavior, Fig.~\ref{fig:time_vs_A} shows the time per training step as a function of the nucleon number $A$ across all calculations in this work. Each data point has been scaled proportionally to the number of GPUs used and inversely to the number of Monte Carlo samples, in order to isolate the intrinsic scaling with system size. For $A \geq 10$, the observed scaling is consistent with an approximate $A^3$ dependence, indicating a subdominant cost of the local-energy calculation. Smaller systems exhibit more favorable scaling, likely due to improved computational efficiency from vectorization and reduced overhead. For larger systems, deviations from ideal scaling emerge, which can be attributed to increased communication requirements (as more GPUs are employed) and the local-energy calculation, which scales as $A^5$, becoming more relevant.

The effect of parallelization is illustrated in Fig.~\ref{fig:log_scaling}, which shows the scaling of the algorithm with the number of GPUs for a representative calculation of $^{40}\mathrm{Ca}$. The time per sample is defined as the average wall-clock time per training step divided by the number of samples in that step, with all results obtained using 16{,}000 samples. Near-ideal scaling is observed for smaller numbers of GPUs, with modest degradation at larger device counts, consistent with communication and synchronization overheads.

\section{Results}
\label{sec:results}

\begin{figure*}[!htb]
\centering
\includegraphics[width=0.97\textwidth]{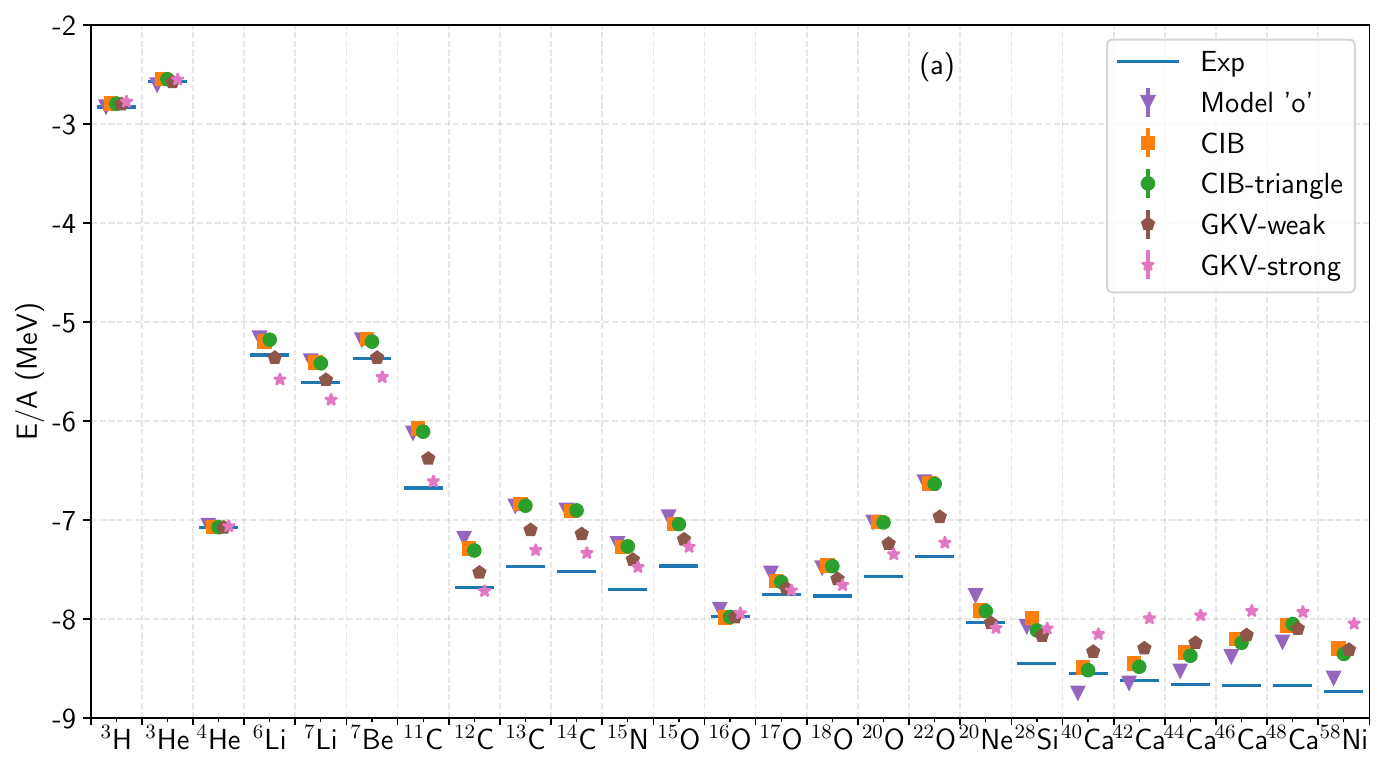}

\vspace{0.1cm}

\includegraphics[width=0.99\textwidth]{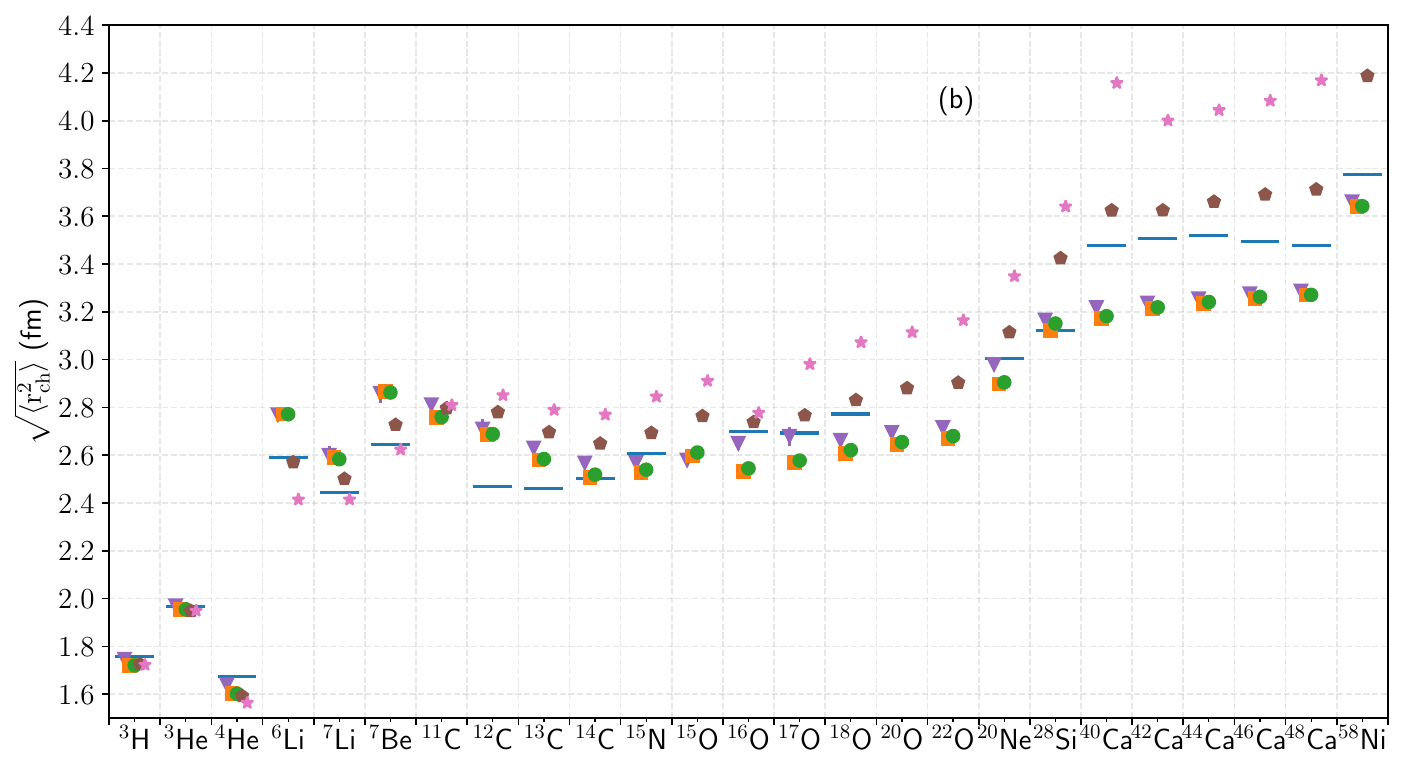}

\caption{Ground-state energy per particle (panel a) and charge radii (panel b) for light and medium-mass nuclei obtained from NQS calculations. Results are shown for the different nuclear Hamiltonians discussed in the text and compared with available experimental data.}
\label{fig:gs_energy_radii}
\end{figure*}

In Fig.~\ref{fig:gs_energy_radii}, we show the ground-state energies per particle and charge radii for nuclei with mass numbers ranging from $A=3$ to $A=58$. Across this mass range, the variational energies reproduce the experimental trend with increasing $A$. As discussed above, model ``o'' provides the baseline for these calculations and achieves an average deviation from experiment of $0.306$ MeV per nucleon, corresponding to about $4\%$ of its average value of 6.918 MeV; see Table~\ref{tab:avg_deviation}. The average discrepancy in the charge radii is also about $4\%$ of its average value of 2.769 fm.

\begin{table}[b]
\centering
\begin{tabular}{c|c}
\hline\hline
Potential & $E(^3\text{He}) - E(^3\text{H})$ [MeV] \\ \hline
Exp.         & 0.764 \\
model ``o''  & 0.666(1) \\
CIB          & 0.744(1) \\
CIB-triangle & 0.744(1) \\
GKV-weak     & 0.671(1) \\
GKV-strong   & 0.672(1) \\
\hline\hline
\end{tabular}
\caption{Energy splitting between the ground states of $^3$He and $^3$H for each interaction, compared with experiment.}
\label{tab:3He3H_diff}
\end{table}

Including CIB terms does not appreciably improve the overall agreement with experiment. In fact, the discrepancy between calculated and measured charge radii increases. This behavior is likely connected to the diagonal filter $P_{ijk}$ of Eq.~\eqref{eq:three_body_proj}, which leads the CIB interactions to underestimate the radii of $^{15}$N and of the oxygen isotopes. On the other hand, CIB terms are important for accurately reproducing the energy splitting between the ground states of $^3$He and $^3$H, shown in Table~\ref{tab:3He3H_diff}. Their inclusion brings the theoretical splitting much closer to experiment for both the linear and triangle $3N$ forces, increasing it from about $0.67$ MeV to about $0.74$ MeV, compared with the experimental value of $0.764$ MeV.

The inclusion of $p$-wave terms has a more pronounced and interaction-dependent impact. The GKV-weak interaction improves the agreement with experiment for both ground-state energies and charge radii, where the average energy deviation is about $3\%$, while the average radius deviation is about $5\%$. These averages are computed over all 25 nuclear species included in this work.

More detailed trends reveal that the GKV interactions generally produce more binding than model ``o'' up to $^{28}$Si. This behavior can be understood from the interplay between the additional $p$-wave attraction and the refitted three-body force. Because $^{4}$He is largely insensitive to the $p$-wave interaction whereas $^{16}$O gains substantial $p$-wave attraction, fitting both nuclei simultaneously requires a significantly stronger and longer-ranged repulsive three-body force; see Table~\ref{tab:3body}. This calibration determines the relative balance between the additional $p$-wave attraction and the repulsive three-body force in $^{16}$O. For the carbon and oxygen isotopes, the additional $p$-wave attraction lowers the energy, while the strengthened three-body force favors a more diffuse spatial distribution. The variational wave function balances both effects simultaneously, becoming more strongly bound through the $p$-wave interaction while expanding enough to reduce the three-body repulsion.

In heavier nuclei, however, the number of interacting triples increases rapidly, amplifying the contribution of the repulsive three-body interaction relative to the additional $p$-wave attraction. As a result, the wave function expands further to reduce this repulsion, leading to larger charge radii and reduced binding. This effect is most pronounced for the GKV-strong interaction, which significantly underbinds the calcium isotopes and $^{58}$Ni while producing charge radii that are consistently too large. By contrast, the GKV-weak interaction is less affected by this mechanism and provides the best overall compromise among the interactions considered here, improving both the ground-state energies and charge radii across much of the nuclear chart, although systematic deviations remain, particularly for neutron-rich isotope chains.

The charge radii show a particularly strong dependence on the choice of Hamiltonian. In contrast to the GKV interactions, model ``o'', CIB, and CIB-triangle tend to overestimate the radii of the lightest nuclei while underestimating those of heavier systems. Overall, the stronger sensitivity of the charge radii to the choice of interaction, compared with the relative stability of the ground-state energies, demonstrates that these observables provide complementary constraints on the underlying balance between attractive and repulsive components of the nuclear interaction.

\begin{table}[!htb]
\centering
\begin{tabular}{c|ccc}
\hline\hline

& $R_n$ [fm]
& $R_p$ [fm]
& $R_{\rm skin}$ [fm] \\
\hline
model ``o'' & 3.4462(5) & 3.1983(4) & 0.2480(5) \\
CIB & 3.3899(8) & 3.1806(9) & 0.2093(11) \\
CIB-triangle & 3.4004(11) & 3.1816(11) & 0.2189(15) \\
GKV-weak & 3.7648(8) & 3.6334(8) & 0.1314(11) \\
GKV-strong & 4.1410(11) & 4.0984(15) & 0.0427(20) \\
\hline
CREX
& --- & --- & 0.121(35) \\
Fayans
& $3.5522^{+0.0021}_{-0.0018}$
& $3.4230^{+0.0025}_{-0.0016}$
& $0.1292^{+0.0016}_{-0.0014}$ \\
\hline\hline
\end{tabular}
\caption{Point-neutron rms radius $R_n$, point-proton rms radius $R_p$, and neutron-skin thickness of $^{48}$Ca. The CREX and Fayans values are taken from Refs.~\cite{CREX:2022kgg} and~\cite{Smith:2026ziu}, respectively.}
\label{tab}
\end{table}

Table~\ref{tab} reports the point-neutron and point-proton rms radii of $^{48}$Ca, together with the neutron-skin thickness $R_{\rm skin}=R_n-R_p$. In addition to the CREX extraction~\cite{CREX:2022kgg}, we include, as a reference, the Fayans EDF values reported in the recent structure--reaction analysis of interaction cross sections along the calcium isotopic chain~\cite{Smith:2026ziu}.

Among the interactions considered here, GKV-weak gives a neutron-skin thickness compatible with CREX and close to the Fayans value. However, this agreement should be interpreted with caution, because its absolute proton and neutron radii are substantially larger than the
Fayans estimates. This is consistent with the charge-radius systematics shown in the lower panel of Fig.~\ref{fig:gs_energy_radii},
where GKV-weak overpredicts the experimental charge radius of $^{48}$Ca. Conversely, model ``o'' and the CIB interactions predict smaller absolute radii, especially for the proton distribution, but overpredict the neutron skin relative to both CREX and the Fayans constraint. GKV-strong provides the clearest counterexample: it  significantly underpredicts the neutron skin while
producing very large absolute radii. These comparisons show that neutron-rich observables are highly sensitive to the details of the Hamiltonian, and that reproducing the neutron-skin thickness alone is not sufficient to validate an interaction; the absolute proton and neutron radii provide essential complementary constraints.

Notably, the Pfaffian-Jastrow NQS ansatz, augmented with MPNN backflow transformations, allows us to describe both open- and closed-shell nuclei within the same framework. To assess this property, Table~\ref{tab:angular_momentum} compares the orbital-angular-momentum expectation values computed from the variational wave functions with the values expected from experimental assignments and simple shell-model configurations. The expected values correspond to $L(L+1)$, in units of $\hbar^2$, and should not be interpreted as direct experimental measurements of $L^2$.

\begin{table}[!b]
\caption{Expected and computed orbital-angular-momentum expectation values for the nuclear ground states. The expected values correspond to $L(L+1)$, inferred from experimental $J^\pi$ assignments together with simple shell-model configurations. The last column gives the values obtained from NQS calculations using the GKV-weak potential. Results from other potentials are similar except for $^{58}$Ni which has significant contamination from nearby excited rotational states.}
\label{tab:angular_momentum}
\begin{tabular}{c|cc}
\hline\hline
Nucleus & Expected $\langle L^2\rangle$ & NQS $\langle L^2\rangle$ \\
\hline
$^{3}$H & 0 & 0.0001(1) \\
$^{3}$He & 0 & 0.0000(0) \\
$^{4}$He & 0 & 0.0000(1) \\
$^{6}$Li & 0 & 0.0001(1) \\
$^{7}$Li & 2 & 2.0012(2) \\
$^{7}$Be & 2 & 2.0013(2) \\
$^{11}$C & 2 & 2.0198(5) \\
$^{12}$C & 0 & 0.0066(2) \\
$^{13}$C & 2 & 2.0692(11) \\
$^{14}$C & 0 & 0.0146(5) \\
$^{15}$N & 2 & 2.0254(5) \\
$^{15}$O & 2 & 2.0334(5) \\
$^{16}$O & 0 & 0.0206(8) \\
$^{17}$O & 6 & 0.0287(5) \\
$^{18}$O & 0 & 0.0262(16) \\
$^{20}$O & 0 & 0.0404(17) \\
$^{22}$O & 0 & 0.0391(8) \\
$^{20}$Ne & 0 & 0.0218(5) \\
$^{28}$Si & 0 & 0.1401(35) \\
$^{40}$Ca & 0 & 0.2686(27) \\
$^{42}$Ca & 0 & 0.3671(30) \\
$^{44}$Ca & 0 & 0.3917(36) \\
$^{46}$Ca & 0 & 0.3638(30) \\
$^{48}$Ca & 0 & 0.3830(28) \\
$^{58}$Ni & 0 & 2.0288(160) \\
\hline\hline
\end{tabular}
\end{table}

For light nuclei, the agreement is generally excellent, indicating that the variational states reproduce the expected angular-momentum structure. For heavier systems, the deviations become more visible. In closed-shell and even-even nuclei, where the expected value is $\langle L^2\rangle=0$, the computed values remain small but become systematically nonzero as the mass number increases. This behavior is consistent with residual admixtures of nearby states, whose density increases in heavier nuclei and which are more difficult to separate within an energy-only optimization framework~\cite{Sun:2024iht}. This issue is especially relevant for $^{58}$Ni, where the various potentials we investigate have widely varying results. Collective and rotational correlations can have a relatively small effect on the total binding energy while still having a significant impact on the shape and angular-momentum content of the wave function.

The most significant deviation is observed in $^{17}$O, for which the expected value $\langle L^2\rangle=6$ is not reproduced. Instead, the NQS value is much closer to zero. In a simple shell-model picture, the ground state of $^{17}$O corresponds to a closed $^{16}$O core plus one valence neutron in the $1d_{5/2}$ orbital, giving $L=2$ and therefore $L^2=L(L+1)=6$. However, $^{17}$O also has a low-lying $1/2^+$ excited state within about $1$ MeV of the ground state, associated predominantly with a $2s_{1/2}$ neutron configuration. The proximity of these two positive-parity states makes the variational optimization particularly sensitive to configuration mixing and to the ordering of the $2s_{1/2}$ and $1d_{5/2}$ configurations.

This behavior likely reflects limitations of the interaction models, in addition to the difficulty of separating low-lying states within an energy-only optimization framework. None of the interactions used in this work contains explicit tensor or spin-orbit operators, which are essential for reproducing the empirical ordering and splitting of the $sd$-shell single-particle levels~\cite{Pieper:1993so}. In particular, the spin-orbit interaction lowers the $1d_{5/2}$ state relative to the $1d_{3/2}$ state, while the $2s_{1/2}$ level is not affected by this splitting. Without these operator structures, the Hamiltonian can favor an incorrect ordering or linear combination of the nearby $2s_{1/2}$ and $1d_{5/2}$ configurations.

To corroborate this interpretation, we performed AFDMC calculations of $^{17}$O with model ``o'', using trial wave functions in which the valence neutron occupies either the $2s_{1/2}$ or the $1d_{5/2}$ orbital. The constrained-path energy obtained with the $2s_{1/2}$ configuration, corresponding to $L^2=0$, is about $1$ MeV lower than that obtained with the $1d_{5/2}$ configuration, corresponding to $L^2=6$. This supports the interpretation that the anomalously small value of $\langle L^2\rangle$ is driven primarily by the Hamiltonian, rather than by a failure of the NQS ansatz alone.

This observation also highlights the importance of future work aimed at targeting states with definite quantum numbers. This could be achieved either by adding penalty terms that constrain the angular-momentum content of the variational state~\cite{Gnech:2023prs}, or by explicitly targeting excited states~\cite{Pfau:2023azx,Zhang:2026iex}. Such developments will be essential for computing transition amplitudes between states with controlled quantum numbers.

\section{Conclusions}
\label{sec:conclusions}
We present, to the best of our knowledge, the first NQS calculations of medium-mass nuclei, reaching up to $^{58}$Ni. The systems considered include very light nuclei such as $^3$H, $^3$He, and $^4$He; clustered nuclei such as $^{12}$C and $^{20}$Ne; oxygen and calcium isotopes; and $^{28}$Si. The computed energies per particle follow the expected systematic behavior with increasing mass number, and the variational results remain close to empirical values across the nuclear chart.

The original essential interaction, model ``o'' of Ref.~\cite{Schiavilla:2021dun}, provides a robust baseline, with an average energy deviation of $0.306$ MeV per nucleon, corresponding to $4.4\%$ of the energy per particle. Adding CIB terms and additional electromagnetic contributions does not significantly improve the global agreement with experiment, although it considerably improves the energy splitting between $^3$H and $^3$He. By contrast, adding weak $p$-wave contributions appreciably reduces the discrepancy with experiment, lowering the average energy deviation to $0.223$ MeV per nucleon, corresponding to $3.2\%$. The strong $p$-wave parametrization instead worsens the agreement in the medium-mass region, owing to the excessively repulsive $3N$ force required to reproduce the binding energy of $^{16}$O. 

Charge radii provide complementary information on nuclear structure that is not contained in ground-state energies alone~\cite{Miyagi:2025lmv}. Together with matter radii, they can reveal extended density distributions characteristic of halo nuclei~\cite{Tanihata:2013jwa}. Along isotopic chains, charge radii are also sensitive to shell closures, deformation, and pairing correlations, including proton superfluidity~\cite{Miller:2019,GarciaRuiz:2019cog,Naito:2022vnz,Yang:2022wbl}.

In the present calculations, charge radii show a stronger dependence on the choice of interaction than the ground-state energies. The baseline model ``o'' produces radii that follow reasonable trends with mass number, with an average deviation of $0.127$ fm. However, it systematically underpredicts the size of the calcium isotopes and of $^{58}$Ni. The CIB interactions exhibit a similar behavior, indicating that CIB terms in the $NN$ interaction have only a minor impact on charge radii at this level.

The GKV interactions show a more pronounced dependence on the strength of the $p$-wave terms. GKV-strong leads to radii that increase too rapidly with system size, despite producing acceptable energies in several nuclei. This behavior is consistent with the strong repulsive $3N$ force required to compensate the additional $p$-wave attraction in $^{16}$O. By contrast, GKV-weak provides the best overall description of charge radii, with an average deviation of $0.138$ fm, corresponding to about $4.7\%$. Interestingly, most measured radii lie between the predictions of model ``o'' and GKV-weak, with the exception of $^{12}$C and $^{13}$C. The discrepancies observed for these highly clustered nuclei suggest that additional higher-order terms in the Hamiltonian, and possibly improved treatment of cluster correlations, may be required.

Given their ability to solve the nuclear quantum many-body problem with high accuracy, NQS provide a promising framework for identifying the essential ingredients of nuclear binding. Rather than fixing the range and strength of the $3N$ force using only $^4$He and $^{16}$O, a natural next step is to perform a global optimization of the Hamiltonian, in the spirit of the UNEDF optimization program for nuclear density functionals~\cite{Kortelainen:2010hv,Kortelainen:2013faa}. Such a fit could include interactions with both $s$- and $p$-wave components, as well as CIB terms, and could simultaneously constrain the $NN$ low-energy constants with scattering data and finite-nucleus observables.

Exploring this enlarged parameter space requires supplementing the transfer-learning procedure used here with emulators~\cite{DiDonna:2025oqf,Li:2026piq}. Once trained on a selected set of NQS calculations, these emulators would provide inexpensive predictions for nuclear observables as functions of the Hamiltonian parameters, enabling global fits and uncertainty quantification along the lines of recent nuclear-structure applications~\cite{Duguet:2023wuh}.

\section*{Acknowledgments}
We thank Chlo\"e Hebborn and Andrew Smith for providing the Fayans radii of $^{48}$Ca and for helpful correspondence. We also thank Robert Wiringa for his valuable comments and suggestions. The present research is supported by the U.S. Department of Energy, Office of Science, Office of Nuclear Physics, under contracts DE-AC02-06CH11357, by the DOE Early Career Research Program, and under the STREAMLINE 2 Collaboration Award. A.~L. also acknowledges support from grant PID2023-147458NB-C21, funded by MCIN/AEI/10.13039/501100011033, and by the European Union. An award of computer time was provided by the U.S. Department of Energy’s (DOE) Innovative and Novel Computational Impact on Theory and Experiment (INCITE) Program. This research used resources from the Argonne Leadership Computing Facility, a U.S. DOE Office of Science user facility at Argonne National Laboratory, which is supported by the Office of Science of the U.S. DOE under Contract No. DE-AC02-06CH11357.

\bibliography{references}

\end{document}